\documentclass[useAMS,usenatbib]{mn2e}

\usepackage{xcolor}
\usepackage{graphicx}
\usepackage{natbib}
\usepackage{grffile}
\definecolor{light-gray}{gray}{0.6}

\usepackage{amssymb}
\usepackage{amsmath}
\usepackage[colorlinks=true, citecolor=blue,linkcolor=blue, urlcolor=red]{hyperref}
\usepackage{multirow}
\usepackage{enumerate}

\newcommand{\Msun}{\mbox{M$_\odot$}}

\title[Galaxy formation at high-redshift]{Galaxy formation in the {\it Planck} Cosmology - III. The high-redshift universe}
\author[Scott J. Clay]{Scott J. Clay$^{1}$\thanks{E-mail:
s.clay@sussex.ac.uk}, Peter A. Thomas$^{1}$, Stephen M. Wilkins$^{1}$ and Bruno M. B. Henriques$^{2}$\\
$^{1}$Astronomy Centre, University of Sussex, Brighton, BN1 9QH, U.K.\\
$^{2}$Max-Planck-Institut f$\ddot{u}$r Astrophysik, Karl-Schwarzschild-Str. 1, D-85741 Garching bei M$\ddot{u}$nchen, Germany}
\begin{document}

\date{Accepted 2015 April 10. Received 2015 April 09; in original form 2015 March 04}

\pagerange{\pageref{firstpage}--\pageref{lastpage}} \pubyear{2015}

\maketitle

\label{firstpage}


\begin{abstract}

We present high-redshift predictions of the
star formation rate distribution function (SFR\,DF), UV luminosity function
(UV\,LF), galactic stellar mass function (GSMF), and specific star-formation
rates (sSFRs) of galaxies from the latest version of the Munich semi-analytic model {\sc
  L-Galaxies}. We find a good fit to both the shape and normalisation of the SFR\,DF at
$z=4-7$, apart from a slight under-prediction at the low SFR end at $z=4$.
Likewise, we find a good fit to the faint number counts for the observed UV\,LF;
at brighter magnitudes our predictions lie below the observations, increasingly
so at higher redshifts.  At all redshifts and magnitudes, the raw (unattenuated)
number counts for the UV\,LF lie above the observations. Because of the good agreement with the SFR we interpret our
under-prediction as an over-estimate of the amount of dust in the model for the
brightest galaxies, especially at high-redshift. While the shape of our GSMF matches that of the observations, we lie between
(conflicting) observations at $z=4-5$, and under-predict at $z=6-7$. The sSFRs of
our model galaxies show the observed trend of increasing normalisation with
redshift, but do not reproduce the observed mass dependence.Overall, we conclude that the latest version of {\sc L-Galaxies}, which is tuned
to match observations at $z\leq3$, does a fair job of reproducing the observed
properties of galaxies at $z\geq4$.  More work needs to be done on understanding
observational bias at high-redshift, and upon the dust model, before strong
conclusions can be drawn on how to interpret remaining discrepancies between the
model and observations.

\end{abstract}


\begin{keywords}
galaxies: evolution -- galaxies: formation -- galaxies: luminosity function, mass function -- galaxies: high--redshift -- ultraviolet: galaxies 
\end{keywords}


\section{Introduction}

With the installation of Wide Field Camera 3 (WFC3) on the {\em Hubble Space
  Telescope (HST)} in 2009 it is now possible to identify statistically useful
and robust samples of star forming galaxies in the early Universe
\citep[$z>4$,][]{Oesch2010,Bouwens2010b,Bunker2010,Wilkins2010,Finkelstein2010,McLure2010,Wilkins2011a,Lorenzoni2011,Bouwens2011,McLure2011,Finkelstein2012b,Lorenzoni2013,McLure2013,Duncan2014,Finkelstein2014}. In
recent years a tremendous effort has been dedicated to quantifying the
photometric and physical properties, such as star formation rates and stellar
masses, of these galaxies. As we continue to dig deeper, with the first sources
now identified at $z \approx 10$ \citep[e.g.][]{Oesch2012a,Ellis2013}, and with
the launch of the {\em James Webb Space Telescope (JWST)} in the next few years,
we will further be able to constrain the physics of galaxy formation and
evolution in this critical epoch of the Universe's history.

Although it lasts less than 0.8\,Gyr, the period of the Universe between $z=7$
and $z=4$ is important to study because it defines an epoch of interesting
galaxy formation and evolution activity. The start of this period marks the end
of the epoch of reionization; galaxies are starting to ramp up their metal and
dust production; and we are finding evidence of the first quasars.  While
astronomy is unique in allowing us to observe the Universe at these early times,
theoretical modelling is required to interpret those observations in terms of an
evolving galaxy population. The rapidly advancing observational constraints on
the physical properties of galaxies in the early Universe provides an
opportunity to further test and refine these galaxy formation models.

The most well studied property of the galaxy population at high-redshift (in
part due to its accessibility) is the rest-frame ultraviolet (UV) luminosity
function (LF). Because of the link between the UV luminosity of galaxies and
their star-formation rates, the observed UV\,LF provides an important constraint
on star-formation activity in the early Universe. While early observational
results were based on only small samples
\citep{Bouwens2008,Bouwens2010a,Bouwens2010b,Bunker2010,Oesch2009,Oesch2010,Ouchi2009,Wilkins2011b,Robertson2010,Dunlop2010,Lorenzoni2011},
we have slowly begun building larger catalogues, first with $200-500$ galaxies
\citep{Finkelstein2010, Bouwens2011, McLure2013}, with the most recent
observations having almost 1000 galaxies at $z\geq7$ \citep{Bouwens2015, Finkelstein2014}.

While the intrinsic UV luminosity is known to be a useful diagnostic of
star-formation activity \citep[e.g.~][]{Wilkins2012}, it is susceptible to even
small amounts of dust ($A_{\rm UV}\approx 10\times E(B-V)$). Direct comparison
of the observed UV luminosity function with predictions from galaxy formation
models is then sensitive to the reliability of the dust model (which has to
account for the creation and destruction of dust, and its effect on the
intrinsic spectral energy distribution).

Whilst challenging, it is observationally possible to constrain the dust
obscuration and thus determine the true (or intrinsic) star formation activity,
even in distant galaxies. Starlight that is absorbed by dust is reprocessed and emitted in
the rest-frame mid/far-IR. Combining the star-formation rate inferred from the
observed UV with that inferred from the mid/far-IR emission then provides a
robust constraint on the total (or intrinsic) star-formation
activity. Observational constraints on the rest-frame mid/far-IR emission in
high-redshift galaxies are, however, challenging due to the significantly lower flux
sensitivity and poorer spatial resolution of facilities operating at these
wavelengths. Thus far there is only a single galaxy
individually detected in the far-IR at $z>6$ \citep{Riechers2013}. This is,
however, likely to rapidly improve with the completion of the {\em Atacama Large
  Millimetre Array (ALMA)}.

One alternative to using far-IR/sub-mm observations is to take advantage of the
relationship between the rest-frame UV continuum slope $\beta$, which is easily
accessible even at $z\sim 10$ (Wilkins et al. {\em submitted}) and the UV
attenuation \citep[first applied by][]{Meurer1999}. The measurement of $\beta$
in high-redshift galaxies has, in recent years, been the focus of intense study
\citep[e.g.][]{Stanway2005,Bouwens2009,Bunker2010,Bouwens2010a,Wilkins2011b,Dunlop2012a,Bouwens2012,Finkelstein2012a,Castellano2012,Rogers2013,Wilkins2013,Bouwens2014}. Measurements
of the UV continuum slope have been used to effectively correct the observed UV
luminosity function and thus determine the star-formation-rate distribution
function \citep[e.g.][]{Smit2012}. It is important to note, however, that this
relation is sensitive to a number of assumptions \citep[see][]{Wilkins2012,
  Wilkins2013} which introduce both systematic biases and increase the scatter
in individual observations.

By combining space (from {\em Hubble}) and ground-based near-IR
observations ($< 2 \mu $m) from the {\em Infra-red Array Camera (IRAC)} aboard
the {\em Spitzer Space Telescope} it is possible to probe the rest-frame to
optical spectral energy distributions (SEDs) of galaxies at high redshift. This
is critical to deriving robust stellar masses and thus the galaxy stellar mass
function (GSMF). The measurement of stellar masses at high-redshift is,
unfortunately, affected by various issues, including: the low sensitivity of the
{\em IRAC} observations; assumptions regarding the star formation and metal
enrichment history of these galaxies; and the effects of strong nebular emission
\citep[e.g.][]{Wilkins2013}. Despite these obstacles, several groups have now
attempted to measure the galaxy SMF in the high-redshift Universe
\citep[e.g.][]{Stark2009, Labbe2010, Gonzalez2011, Yan2012, Duncan2014}
permitting a direct comparison with galaxy formation models.

The Munich semi-analytic model of galaxy formation \citep[latest
  version][]{Henriques2014}, also known as {\sc L-Galaxies}, has had a lot of
success over the past decade in predicting various properties of galaxies, such
as the stellar-mass and luminosity functions both in the local Universe and out
to redshift $z=3$ \citep{Henriques2013}.  In this paper we extend these
predictions to higher redshift without altering any of the model parameters
(except to modify the redshift-dependence of the dust model, as described in
\S2.2.1 below). In that sense, the results presented here may be considered
predictions of the model.

\begin{figure*}
\begin{center}
\includegraphics[width=0.8\textwidth]{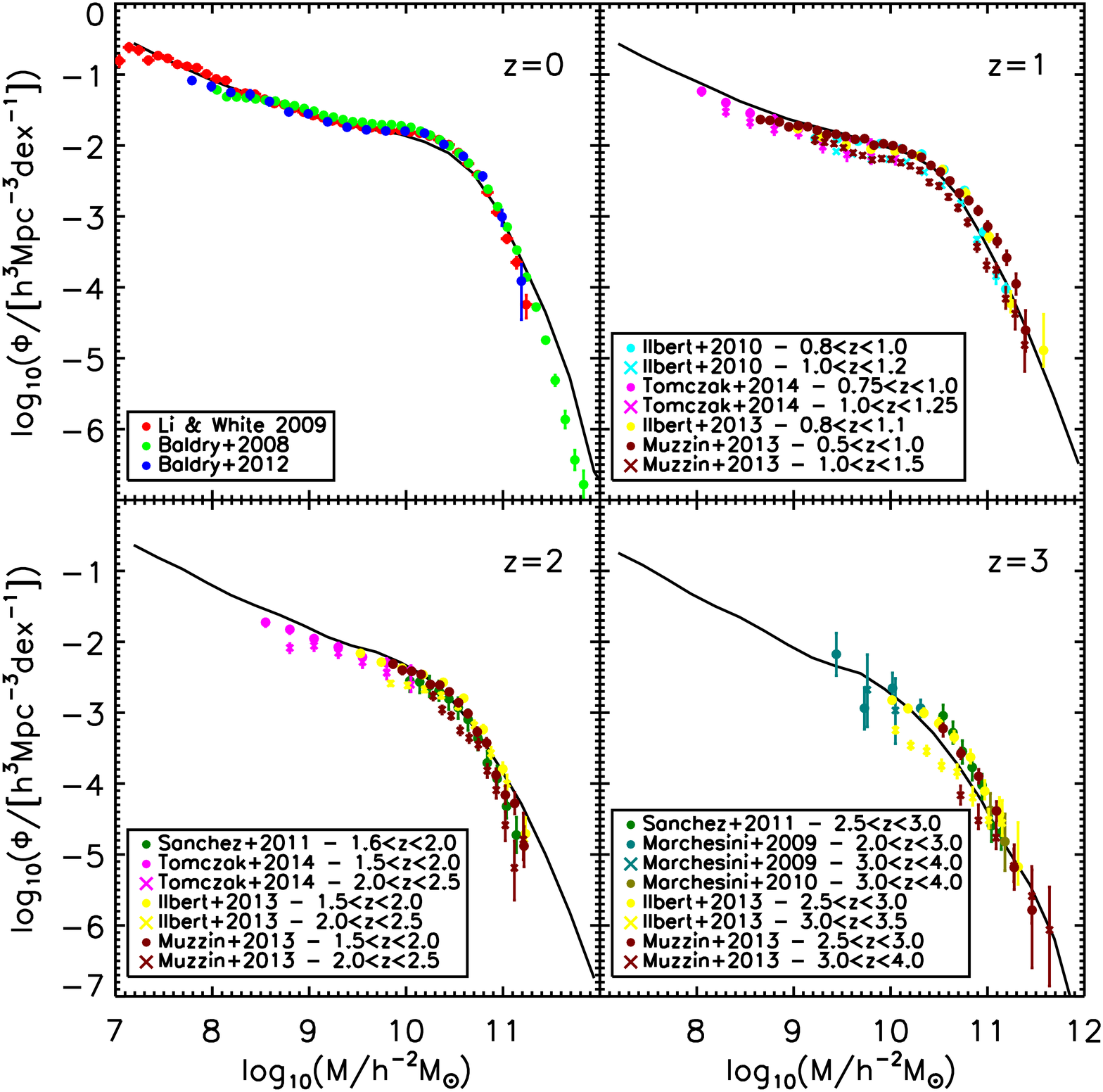}
\centering
\caption{Predicted stellar mass functions at redshift $z \approx 0$ (top left);
  $z \approx 1$ (top right); $z \approx 2$ (lower left) and $z \approx 3$ (lower
  right). Solid black lines show the stellar mass functions predicted by our model. This figure
  is reproduced from \protect\cite{Henriques2014} Figures 2 and A1, and we direct the reader there
  for a more detailed description. We include it here to highlight how well the model works at
  lower redshifts in predicting key observables such as the SMF. Observations are taken from several 
  surveys; SDSS \protect\citep{Baldry2008,Li2009} and GAMA \protect\citep{Baldry2012} at z=0; and \protect\cite{Marchesini2009}, Spitzer-COSMOS \protect\citep{Ilbert2010}, NEWFIRM \protect\citep{Marchesini2010}, COSMOS \protect\citep{Sanchez2011}, ULTRAVISTA \protect\citep{Muzzin2013, Ilbert2013} and ZFOURGE \protect\citep{Tomczak2014} at higher redshifts.}
\label{fig:lowzsmf}
\end{center}
\end{figure*}

This paper is structured as follows: in Section 2 we describe the relevant parts
of our semi-analytical model, highlighting the changes in the latest version; in
Section 3 we discuss our high redshift predictions for the star-formation-rate
distribution function and UV luminosity function, followed by the stellar mass
function in Section 4. In Section 5 we discuss the relationship between the
specific star formation rate and the stellar mass and in Section 6 we give a
brief overview of the evolution of these properties at high-redshift. We
conclude our work in Section 7.  Throughout this paper we adopt a Chabrier
initial mass function \citep{Chabrier2003} and use the latest {\it Planck}
cosmology \citep{PlanckCollaboration2014}.  Number densities are presented per
co-moving volume, ($h^{-1}$\,Mpc)$^3$.

This paper makes use of detailed predictions from the new model of {\sc L-Galaxies} outlined in \cite{Henriques2014}, which have been made
publicly available.\footnote{http://gavo.mpa-garching.mpg.de/MyMillennium/} The binned data used to make the plots in this paper have also been made available online.\footnote{http://astronomy.sussex.ac.uk/$\sim$sc558/}


\section[]{The Model}
\subsection{{\sc L-Galaxies}}
Semi-analytic models (SAMs) provide a relatively inexpensive method of
self-consistently evolving the baryonic components associated with
dark matter merger trees, derived from $N$-body simulations or
Press-Schechter calculations.  The term semi-analytic comes from the
use of coupled differential equations (rather than numerical
calculations), to follow the evolution of galaxy formation physics
determining the properties of gas and stars.  Physics commonly found
in most SAMs include descriptions of: (1) primordial infall and the
impact of an ionizing UV background; (2) Radiative cooling of the gas;
(3) Star Formation recipes; (4) Metal enrichment; (5) Super-massive
black hole growth; (6) Supernovae and AGN feedback processes; (7) The
impact of environment and mergers including galaxy morphologies and
quenching.

The Munich SAM, or {\sc L-Galaxies}, \citep{Springel2001,DeLucia2004,
  Springel2005, Croton2006,DeLucia2007, Guo2011, Guo2013,Henriques2013} has been
developed over the years to include most of the relevant processes that affect
galaxy evolution.  In this work we use its latest version, \cite{Henriques2014},
and direct the reader to the appendix of that paper for a detailed description
of the model.  Of most relevance to this paper are the adoption of the {\sc
  Planck} year 1 cosmology and a modified gas-to-dust relation, partly motivated
by the work presented in this paper. The model parameters were constrained
using the abundance and passive fractions of galaxies at $z\leq3$; and the model has
successfully reproduced key observables at these redshifts, such as the luminosity
and stellar mass functions. We highlight this fact in Figure~\ref{fig:lowzsmf}, which is
a reproduction of the SMF at $z\in\{0,1,2,3\}$ from \cite{Henriques2014}. We direct the reader
to Figures 2 and A1 and the related text of that paper for a more detailed discussion, but we highlight
how well the model can explain the observed evolution in the SMF at these redshifts, over the mass
range constrained by observers.

\subsection{Dust Extinction Model}
\label{sec:msamdust}

Actively star-forming galaxies are known to be rich in dust. This can
have a dramatic effect on their emitted spectrum since dust
significantly absorbs optical/UV light while having a much milder
effect at longer wavelengths. As a result, dust-dominated galaxies
will generally have red colours even if they are strongly
star-forming.  For that reason, we summarise the dust model of
\citet{Henriques2014} here: a fuller description can be found in Section~1.14 of
the supplementary material in that paper.

We considering dust extinction separately for the diffuse interstellar medium
(ISM) and for the molecular birth clouds (BC) within which stars form. The
optical depth of dust as a function of wavelength is computed separately for
each component and then combined as described below.  We do not at present
attempt to compute the detailed properties of the dust particles or the
re-emission of the absorbed light.

\subsubsection{Extinction by the ISM}
\label{sec:msamismdust}
The optical depth of
diffuse dust in galactic disks is assumed to vary with wavelength as
\begin{align} \label{eq:msamextinctionism}
\nonumber \tau_{\lambda}^{ISM}= &
(1+z)^{-1}\left(\frac{A_{\lambda}}{A_\mathrm{V}}\right)_{Z_{\odot}}
\left(\frac{Z_{\rm{gas}}}{Z_{\odot}}\right)^s \\
&  \times \left(\frac{\langle N_H\rangle}{2.1
\times10^{21}{\rm{atoms}}\,{\rm{cm}}^{-2}}\right),
\end{align}
where
\begin{equation} \label{eq:msamhcolumndensity}
\langle N_H\rangle=\frac{M_{\rm{cold}}}{1.4\,m_p\pi (a
R_{\rm{gas,d}})^2}
\end{equation}
is the mean column density of hydrogen.
Here $R_{\rm{gas,d}}$ is the cold gas disk scale-length, $1.4$
accounts for the presence of helium and $a=1.68$
in order for $\langle N_H\rangle$ to represent the mass-weighted
average column density of an exponential disk.  Following the results
in \citet{Guiderdoni1987}, the extinction curve in
eq.~(\ref{eq:msamextinctionism}) depends on the gas metallicity and is
based on an interpolation between the Solar Neighbourhood and the
Large and Small Magellanic Clouds: $s=1.35$ for $\lambda<2000$ \AA
$\:$ and $s=1.6$ for $\lambda>2000$ \AA. The extinction curve for
solar metallicity, $(A_{\lambda}/A_\mathrm{V})_{Z_{\odot}}$, is taken from
\citet{Mathis1983}.

The redshift dependence in eq.~(\ref{eq:msamextinctionism}) is significantly
stronger than in previous versions of our model ($(1+z)^{-0.5}$ in
\citet{Kitzbichler2007} and $(1+z)^{-0.4}$ in \citet{Guo2009}). The dependence
implies that for the same amount of cold gas and the same metal abundance, there
is less dust at high redshift. The motivation comes both from observations
\citep{Steidel2004, Quadri2008} and from the assumption that dust is produced by
relatively long-lived stars. However, it may also be that this redshift
dependence has to be introduced as a phenomenological compensation for the
excessively early build-up of the metal content in model galaxies. In practice
it has been included simply to give an approximate match to the low extinctions
of high-redshift galaxies as inferred from their observed UV slopes
\citep{Bouwens2012}, and to the UV luminosity function, as described below.

\subsubsection{Extinction by molecular birth clouds}
\label{sec:msammoleculardust}
This second source of extinction affects only young stars that are still
embedded in their molecular birth clouds, for which we assume a lifetime of 10\,Myr.
The relevant optical depth is taken to be
\begin{equation} \label{eq:msamextinctionclouds}
 \tau_{\lambda}^{BC}=\tau_{\lambda}^{\mathrm{ISM}}\left(\frac{1}{\mu}-1\right)
 \left(\frac{\lambda}{5500\mathrm{\AA}}\right)^{-0.7},
\end{equation}
where $\mu$ is given by a random Gaussian deviate with mean 0.3
and standard deviation 0.2, truncated at 0.1 and 1.

\subsubsection{Overall extinction curve}
\label{sec:msamdustgeometry}
In order to get the final overall extinction, every galaxy is
assigned an inclination, $\theta$, given by the angle between the disk angular
momentum and the $z$-direction of the simulation box, and a ``slab''
geometry is assumed for the dust in the diffuse ISM.  For sources that are
uniformly distributed within the disk then the mean absorption coefficient is
\begin{equation} \label{eq:msamextinctionlambda}
A_{\lambda}^\mathrm{ISM}=-2.5\log_{10}\left(\frac{1-\exp^{-\tau_{\lambda}^\mathrm{ISM}\sec{\theta}}}
{\tau_{\lambda}^\mathrm{ISM}\sec{\theta}}\right), 
\end{equation}
Emission from young stars embedded within birth clouds is
subject to an additional extinction of
\begin{equation} \label{eq:msamextinctionlambda2}
A_{\lambda}^\mathrm{BC}=-2.5\log_{10}\left(\exp^{-\tau_{\lambda}^\mathrm{BC}}\right).
\end{equation}

The standard {\sc L-Galaxies} output does not attempt to model the attenuation
of light by the intergalactic medium. However, this is done in post-processing
for the lightcones published in the Millennium Run Observatory\footnote{The
  Millennium Run Observatory, or MRObs, allows you to observe our semi-analytic
  galaxy formation model through the use of `virtual
  telescopes'.}\citep{Overzier2013}. In this paper, however, we neglect
intergalactic attenuation.


\section{Recent Star Formation}
\label{sec:sfr}

In this section we investigate the star formation rate, and the related UV
luminosity function, at redshifts $z\in\{4,5,6,7\}$.

\begin{figure*}
\begin{center}
\includegraphics[width=0.8\textwidth]{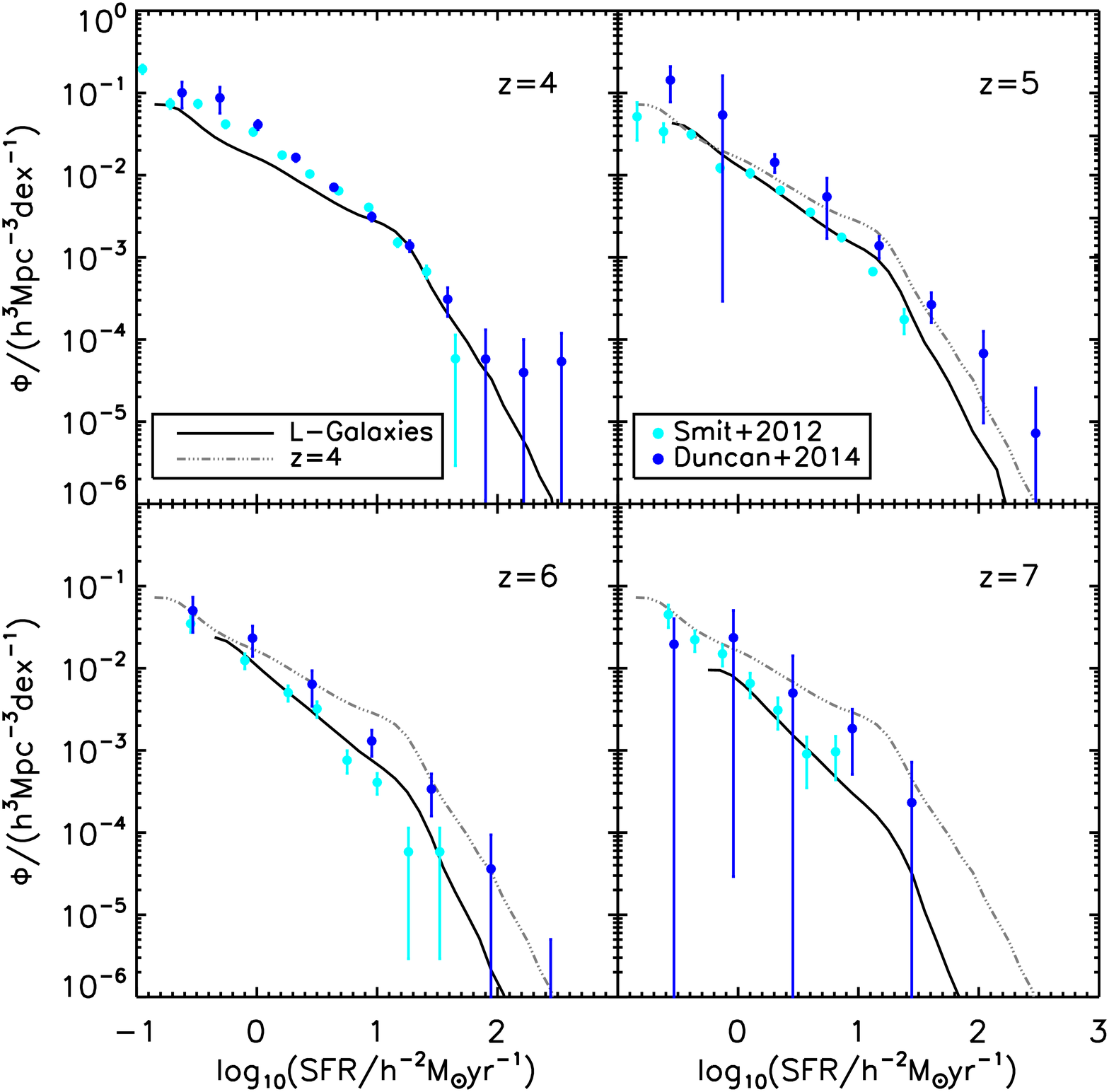}
\centering
\caption{Predicted star formation rate distribution functions at redshift $z
  \approx 4$ (top left); $z \approx 5$ (top right); $z \approx 6$ (lower left)
  and $z \approx 7$ (lower right). In each instance we use the closest available
  snapshot from our {\sc L-Galaxies} run of z=3.95, 5.03, 5.82 and 6.97
  respectively. Solid black lines show the star formation rate distribution
  function predicted by our model. Our $z=4$ star formation rate distribution
  function is repeated at higher redshifts as a grey dot-dash line for
  comparison. Observations are taken from \protect\cite{Smit2012}, converted to
  a Chabrier IMF, and \protect\cite{Duncan2014}.}
\label{fig:sfr}
\end{center}
\end{figure*}

Figure~\ref{fig:sfr} shows the star-formation-rate distribution function
(SFR\,DF) as predicted by our model alongside measurements from \cite{Smit2012}
(converted to our fiducial Chabrier IMF) and \cite{Duncan2014}.

Comparing with the \cite{Smit2012} measurements at redshifts $z\approx 5-7$ we
find generally good agreement.  At these redshifts, the \cite{Duncan2014}
measurements are generally higher than both the model and the \cite{Smit2012}
results. This is particularly true for the most massive galaxies, though we note
that the quoted observational uncertainties can be very large.

At $z\approx4$, however, our model under-predicts the number of galaxies for
$\log_{10}($SFR/$h^{-2}\Msun)<1$ when compared to both sets of observations
(which are consistent with one another at this redshift).  The cause of the
discrepancy is unclear, though may be a consequence of our model
under-estimating the contribution to the SFR from merger-driven activity.

\subsection{The UV Luminosity Function}
\label{sec:uvlf}

\begin{figure*}
\begin{center}
\includegraphics[width=0.8\textwidth]{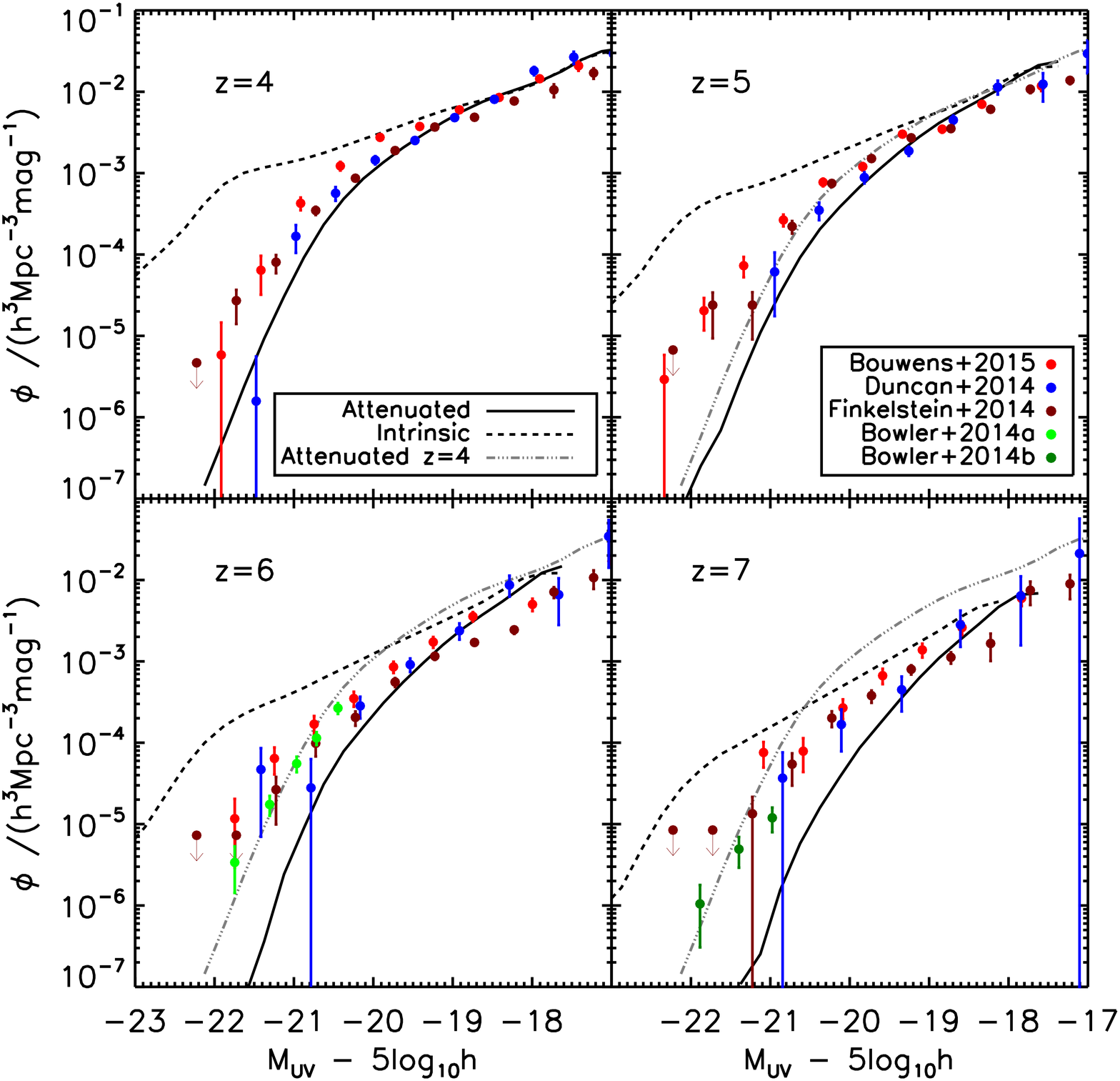}
\centering
\caption{Predicted rest--frame (1500\,\AA) UV luminosity functions at redshift
  $z \approx 4$ (top left); $z \approx 5$ (top right); $z \approx 6$ (lower
  left) and $z \approx 7$ (lower right). In each instance we use the closest
  snapshot available to use from our {\sc L-Galaxies} run of z=3.95, 5.03, 5.82
  and 6.97 respectively. Solid black lines shows the {\sc L-Galaxies} prediction
  for the attenuated UV luminosity function using the dust extinction model
  outlined in \S2.2.1. The dashed black line is the {\sc L-Galaxies} prediction
  of the intrinsic UV luminosity function, with no dust model applied. Our $z=4$
  attenuated UV luminosity function is repeated at higher redshifts as a grey
  dot-dash line for comparison. Observations are taken from
  \protect\cite{Bouwens2015}, \protect\cite{Duncan2014} and
  \protect\cite{Finkelstein2014}, and at high mass from
  \protect\cite{Bowler2014a} ($z=6$) and \protect\cite{Bowler2014b} ($z=7$).}
\label{fig:uvlf}
\end{center}
\end{figure*}

We present the UV luminosity function predicted by our model in Figure
\ref{fig:uvlf} alongside recent observational estimations at high-redshift
\citep{Bouwens2015,Duncan2014,Finkelstein2014,Bowler2014a,Bowler2014b}. The
solid black line shows our prediction for the attenuated UV luminosity function;
the attenuated UV LF at $z=4$ is also shown on
subsequent plots for comparison. The dashed line shows our intrinsic UV
luminosity function, with no dust model being applied. 

We find a good fit to the faint number counts: $M_\mathrm{UV}>-20$ for $z=4-6$
and $M_\mathrm{UV}>-19$ for $z=7$.  At brighter absolute magnitudes, the model
counts fall below the observed ones.  Note, however, that the raw counts, before
dust attenuation, lie above the observations.  Given that we saw a good fit in
Section~\ref{sec:sfr} between predicted and observed SFRs, then this points to a
difference in the dust model between the two.

\begin{figure*}
\begin{center}
\includegraphics[width=0.8\textwidth]{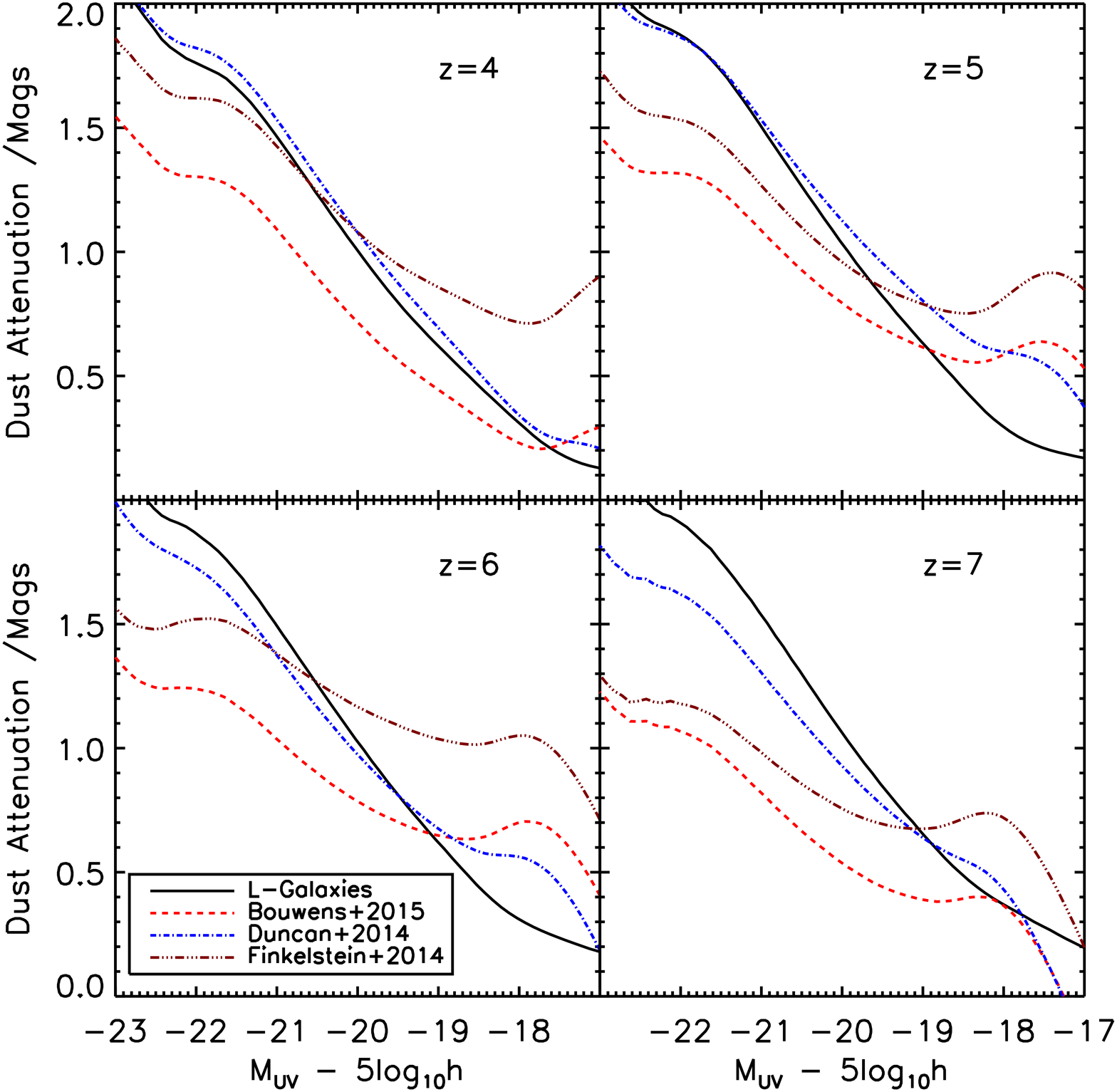}
\centering
\caption{This figure shows the amount of dust attenuation required to move our
  intrinsic UV luminosity function (the dashed, black lines in
  Figure~\protect\ref{fig:uvlf}) to match different observational data sets
  \protect\citep{Bouwens2015, Duncan2014, Finkelstein2014}, as a function of
  unattenuated absolute UV magnitude.  The solid, black line shows the attenuation
  built into the {\sc L-Galaxies} model as described in Section~\ref{sec:msamdust}.}
\label{fig:dustatt}
\end{center}
\end{figure*}

To better understand this, we quantify in Figure~\ref{fig:dustatt} the
attenuation required (as a function of the intrinsic UV absolute magnitude) to
reconcile the raw {\sc L-Galaxies} data with observations.  We do this by
comparing observed, $M_{\Phi,\mathrm{obs}}$, and intrinsic,
$M_{\Phi,\mathrm{int}}$, absolute magnitudes below which we achieve a particular
cumulative number density, $\Phi$ of galaxies:
\begin{equation}
\Phi = \int_{-\infty}^{M_\Phi}\phi\,\mathrm{d}M,
\label{eq:Phi}
\end{equation}
where $\phi$ is the usual differential number density of galaxies.
The attenuation is then $A_\mathrm{UV}=M_{\Phi,\mathrm{obs}}-M_{\Phi,\mathrm{int}}$.

The dust attenuation required to match the observation (as a function of the
intrinsic absolute magnitude) is shown in Figure~\ref{fig:dustatt}.  The black,
 solid line shows the attenuation currently implemented in {\sc L-galaxies}, as
 described in Section~\ref{sec:msamdust}.

As expected, the built-in attenuation matches that from the \citet{Duncan2014}
data fairly well. The other data sets show a shallower slope: the attenuation is
  reasonable, perhaps even under-estimated in the faintest galaxies, but is
  strongly over-estimated in the brightest galaxies and increasingly so at high
  redshift.

It is important to stress that while we are presenting the results for all the objects within
our simulation, observational samples  \citep[such as those employed by][]{Bouwens2015, Duncan2014, Finkelstein2014}, are 
biased and may not truly capture the full galaxy population at these redshifts. Indeed,
a defining characteristic of the Lyman break technique, which is regularly used to 
identify galaxies in the high redshift universe, is that it preferentially selects blue
rest-frame UV bright sources, i.e star forming galaxies with low UV dust 
attenuation $(\rm{A_{UV}}<2$). Very dusty galaxies, or those with little to no star formation 
would then be missed in typical Lyman break galaxy searches \citep[e.g. HFLS3, a very dusty intensely star forming galaxy at $z\approx 6.3$][]{Riechers2013}. The degree to which 
this is a concern at high-redshift is difficult to assess, largely due to the lack
of sensitive far-IR and sub-mm imaging which is critical to identify heavily 
obscured systems.

Given the current observational uncertainties, we conclude that the simple,
empirical dust extinction model currently built into {\sc L-galaxies} does a
reasonable job, although it could be refined to match particular data sets if
required.  
In the future, we intend to implement a more physically-motivated dust model:
we note that the current model has prompt recycling, and this could be
an issue at these early times when the age of the Universe is just 1.5\,Gyr at
$z=4$ and less than 1\,Gyr for $z>6$.  A delayed chemical enrichment model has
been implemented in {\sc L-Galaxies} by \citet{Yates2013} and we intend to incorporate
that into the \citet{Henriques2014} model in future work.


\section{Galaxy Stellar Mass Function}

\begin{figure*}
\begin{center}
\includegraphics[width=0.8\textwidth]{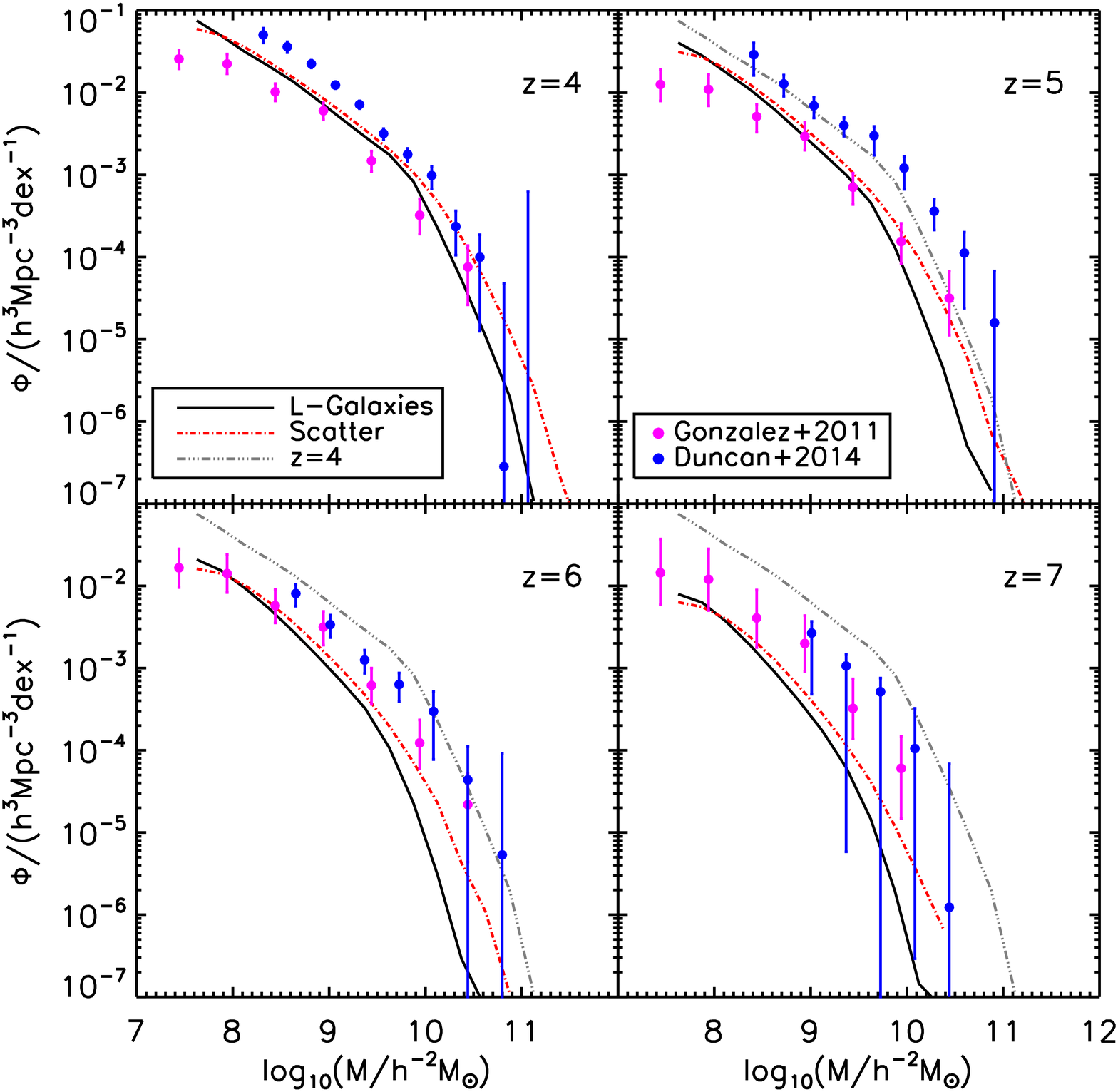}
\centering
\caption{Predicted stellar mass functions at redshift $z \approx 4$ (top left);
  $z \approx 5$ (top right); $z \approx 6$ (lower left) and $z \approx 7$ (lower
  right). In each instance we use the closest snapshot available to use from our
  {\sc L-Galaxies} run of z=3.95, 5.03, 5.82 and 6.97 respectively. Solid black
  lines show the stellar mass functions predicted by our model.  To indicate the
  possible effect of uncertainties in the observational stellar mass
  determinations, we also show as a red dot-dash line the stellar mass function
  convolved with a gaussian of standard deviation 0.3\,dex. Our
  $z=4$ stellar mass function is repeated at higher redshifts as a grey dot-dash
  line for comparison. Observations are taken from \protect\cite{Gonzalez2011},
  converted to a Chabrier IMF, and \protect\cite{Duncan2014}.}
\label{fig:smf}
\end{center}
\end{figure*}

We present the Galaxy Stellar Mass Function (GSMF) at $z\in\{4,5,6,7\}$
predicted by our model in Figure \ref{fig:smf} alongside recent observational
estimates at high-redshift from \cite{Gonzalez2011} and \cite{Duncan2014}.

It is important to first note that the observationally-derived mass functions presented in Figure \ref{fig:smf} are inconsistent with each other at $z\sim 4-5$. One possible source (see \cite{Duncan2014} for a wider discussion) of this discrepancy is the effect of nebular emission which was included in \cite{Duncan2014} but not in \cite{Gonzalez2011}. Galaxies in the high-redshift Universe are expected \citep{Wilkins2013} and inferred \citep[e.g.][]{Smit2014} to exhibit strong nebular emission which can strongly affect the measured stellar mass-to-light ratios and thus masses \citep{Wilkins2013}. The accuracy/precision of stellar mass estimates are also affected by the lower sensitivity and angular resolution of the {\em Spitzer}/IRAC imaging.

Given the above observational uncertainties, it is gratifying that the model
predictions split the two observational measurements at $z=4$.  There is a hint
that the change in slope at the ``knee'' of the mass-function
($M_\mathrm{knee}\approx3\times10^9h^{-2}$\,\Msun) may be sharper in the model than the
observations, but the observational error bars are growing at this point and so
it is hard to draw firm conclusions.  As we move to higher redshifts, however,
the model predictions and the observations gradually diverge as follows: (i) the
normalisation at $M_\mathrm{knee}$ declines more rapidly with increasing
redshift in the models than in the observations; (ii) the slope of the mass
function above the knee is steeper in the models than in the observations.

\begin{figure*}
\begin{center}
\includegraphics[width=0.8\textwidth]{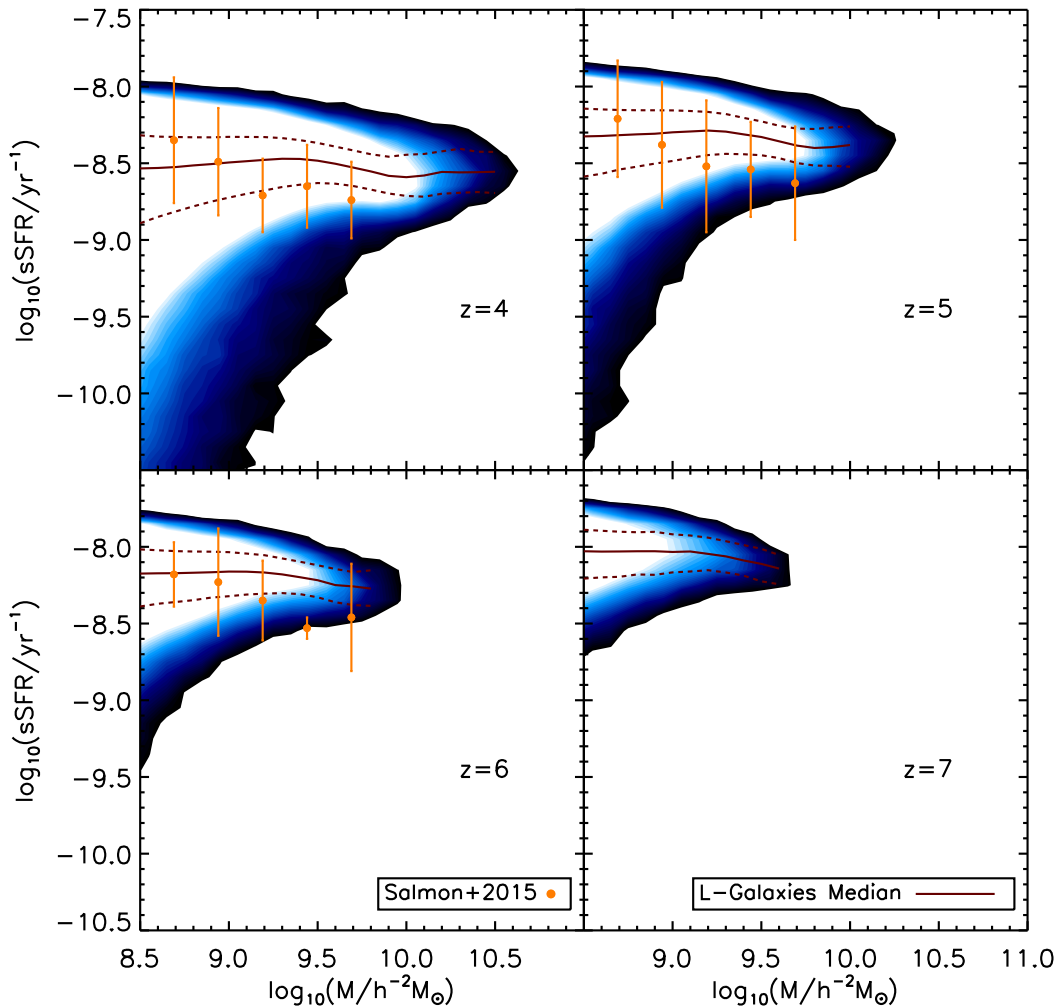}
\centering
\caption{Predicted specific star formation rates ($sSFR = SFR/M$) at redshift $z
  \approx 4$ (top left); $z \approx 5$ (top right); $z \approx 6$ (lower left)
  and $z \approx 7$ (lower right). In each instance we use the closest snapshot
  available to use from our {\sc L-Galaxies} run of z=3.95, 5.03, 5.82 and 6.97
  respectively. The histogram density plot represents the {\sc L-Galaxies}
  galaxy population, with white representing the most dense, and blue
  representing the least. The solid line shows the {\sc L-Galaxies} median
  values, and the dashed lines show the 0.16 and 0.84 percentiles. The observations are taken from \protect\cite{salmon2015}, the points denotes the median while the error bars reflect the scatter in the observed values (not the uncertainty on the median).}
\label{fig:ssfr}
\end{center}
\end{figure*}

The exact cause of these discrepancies is difficult to assess. One possibility
is that it reflects a deficiency in the model; on the other hand it may reflect
a systematic bias in the observations. This has been discussed at low redshift ($z=0-3$)
in Appendix~C of \citet{Henriques2013}. It seems probable that the
uncertainties on the individual stellar masses could have been underestimated, and
that can strongly boost the inferred number of galaxies in regions where
where the mass function is particularly steep.  As an example of the possible
magnitude of this effect, we show in Figure~\ref{fig:smf} the result of
convolving with a gaussian of standard deviation 0.3\,dex, similar to that
required at low redshift.  This largely reconciles the observed and predicted
slopes of the mass function, but the normalisation remains too low at $z=7$. Understanding the source of this discrepancy is a focus of an additional work in progress (Wilkins et al. {\em in-prep}). 

Recent hydrodynamic simulations, particularly {\sc Illustris} \citep{Vogelsberger2014} and {\sc Eagle} \citep{Schaye2015}, have begun making predictions of observables at high redshift \citep{Genel2014, Furlong2014}. Like {\sc L-Galaxies}, both {\sc Illustris} and {\sc Eagle} make predictions at high redshift by only using observational constraints at lower redshift. Both simulations are similar to ours in the prediction of the GSMF at $z=6-7$ in that we all under predict the abundance of high mass galaxies ($>10^9 \Msun$) at these redshifts, although {\sc Eagle} better match the observations at $z=5$ across the entire mass range. Whilst both {\sc L-Galaxies} and {\sc Eagle} match a similar shape to the observations, particularly finding good agreement with the slope and abundance for low mass galaxies, {\sc Illustris} predicts a slope that steepens with increasing redshift faster than what is observed, and over predicts the abundance of low mass galaxies at all redshifts.


\section{Specific Star Formation Rate}

\begin{figure*}
\begin{center}
\includegraphics[width=\textwidth]{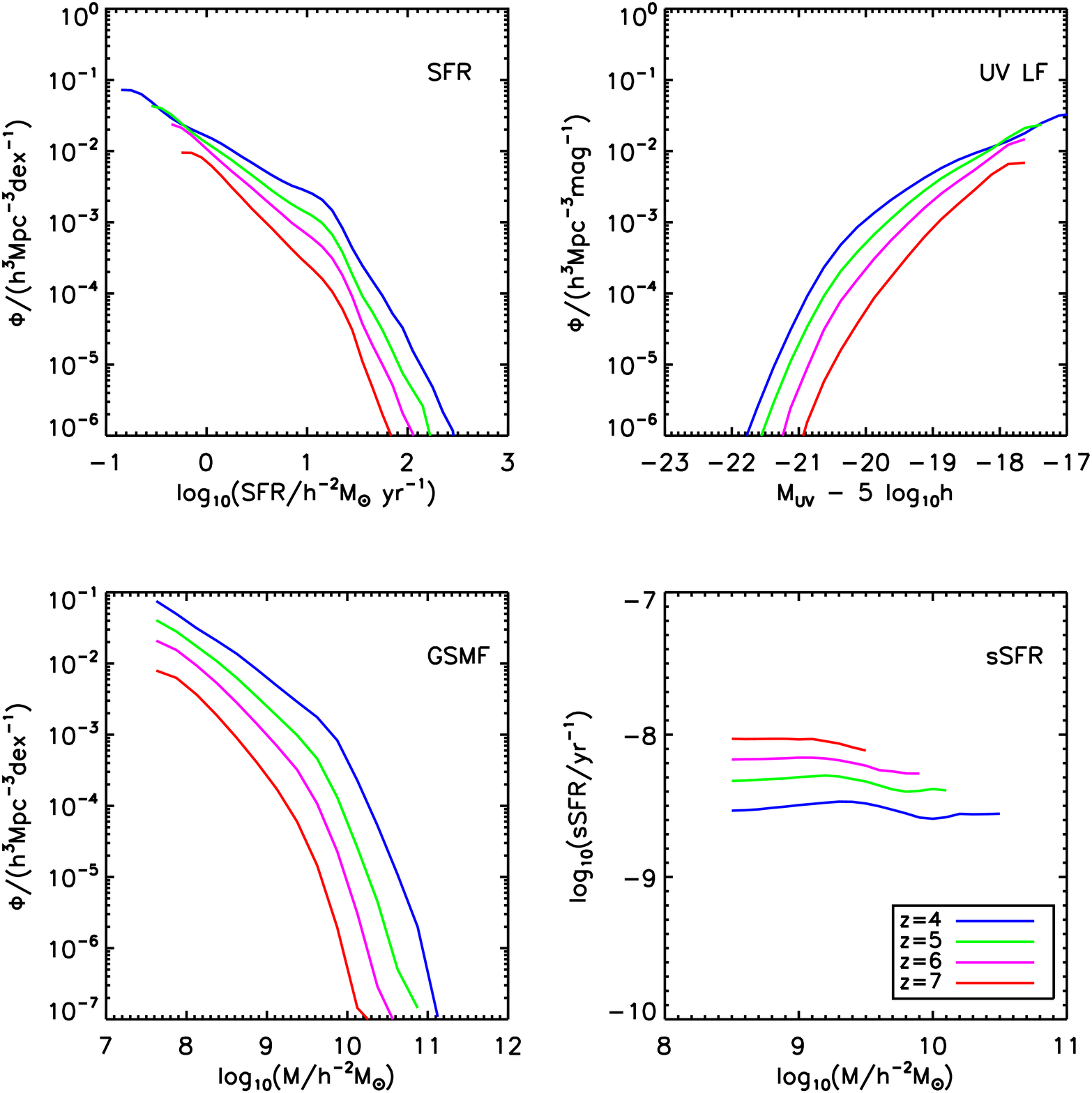}
\centering
\caption{Plot to show the evolution of the stellar mass (top left); star
  formation rate (top right); UV luminosity function (lower left); and specific
  star formation rate (lower right) in the redshift range $z= 4 - 7$. In each
  instance we use the closest snapshot available to use from our {\sc
    L-Galaxies} run of z=3.95, 5.03, 5.82 and 6.97 respectively.}
\label{fig:evolution}
\end{center}
\end{figure*}

The specific star formation rate (sSFR) is a measure of how quickly a galaxy is
forming its stars. We present the sSFRs at $z\in\{4,5,6,7\}$ of our galaxy
population predicted by our model in Figure \ref{fig:ssfr} alongside recent
observational measurements from \cite{salmon2015}. We represent the sSFR of
individual galaxies by a 2D histogram; the solid line shows the median value
predicted by our model, averaged over bins of 100 or more galaxies.

The observations are consistent with our model, particularly for galaxies of mass $M\approx
10^9$\,\Msun, across all redshifts. However the observations show a decline in
the sSFR with increasing galactic stellar mass and we do not identify the same
trend. Instead, all galaxies in our model have roughly the same level of
activity, regardless of galactic stellar mass.  This discrepancy is not
surprising: given that the models match the observed SFR but under-predict the
stellar masses of the largest galaxies, then we would expect this result.

The question remains as to whether the observations or the model is at fault, or
a combination of both.  We could boost AGN feedback in the most massive galaxies
in the model, but this would then reduce the bright end of the UV\,LF.
Alternatively, as hinted in the previous section, the inferred masses of the
highest-mass galaxies may have been boosted by observational scatter.  At lower
masses, there will be an observational bias towards the brightest galaxies, and
so the median sSFR may be over-estimated.


\section{Evolution}
To make it easier to see how the properties of our model galaxies change with
time, we extract the model predictions from each of the four redshifts shown in
Figures~\ref{fig:sfr}, \ref{fig:uvlf}, \ref{fig:smf} \& \ref{fig:ssfr} and
display them in single panels in Figure~\ref{fig:evolution}.

Concentrating first on the SFR (upper-left panel), we see the the knee of the
distribution remains relatively unchanged, at about $20h^{-2}$\Msun\,yr$^{-1}$
over this period.  However, the normalisation of the relation grows and the
slope decreases, such that the number comoving number density of galaxies with
star-formation rates of $0.3h^{-2}$\Msun\,yr$^{-1}$ is approxiately constant,
while that of higher star-formation rates in excess of
$100h^{-2}$\Msun\,yr$^{-1}$ grows by several orders of magnitude.  As might be
expected, a similar, but less pronounced, trend is seen in the UV\,LF, although
the knee of the distribution is harder to discern.

In constrast to the SFR\,DF, the galactic SM\,DF shows only a slight reduction
in slope from $z=7$ to $z=4$.  Consequently, the comoving number density of
low-mass galaxies increases by about 1\,dex over this time.  This is reflected
in the specfic star-formation rate, which reduces by a factor of about 0.5\,dex
in the  same period (as the sSFR is approximately independent of mass, this
conclusion holds for individual galaxis, not just the population).\footnote{The
  age of the Universe roughly doubles over this period; thus the sSFR measured
  in terms of this age shows much less variation and even at $z=4$ is sufficient
  to double the mass of a galaxy in less than a quarter of the age at that time.}


\section{Conclusions}
We have presented the latest high-redshift observational predictions of the
star-formation-rate distribution function (SFR\,DF); UV luminosity function
(UV\,LF); galactic stellar mass function (GSMF) and specific star-formation
rates (sSFRs) of galaxies from the latest version of the {\sc L-Galaxies}
semi-analytic model \citep{Henriques2014}. Our conclusions are as follows:
\begin{enumerate}[(i)]
  \item We find a good fit to both the shape and normalization of the observed
    SFR\,DF at $z=4-7$ (Figure~\ref{fig:sfr}), apart from a slight
    under-prediction at the low SFR end at $z=4$, possibly caused by a lack of
    SFR contribution from merger-driven activity in our model.
  \item We find a good fit to the faint number counts for the observed UV\,LF
    (Figure~\ref{fig:uvlf}).  At brighter magnitudes, our predictions lie below
    the observations, increasingly so at higher redshifts.
  \item At all redshifts and magnitudes, the raw (unattenuated) number counts for
    the UV\,LF lie above the observations, and so we interpret our
    under-prediction as an over-estimate of the amount of dust in the model for
    the brightest galaxies, especially at high-redshift (Figure~\ref{fig:dustatt}).
  \item While the shape of our SMF matches that of the observations, we lie
    between the observations at $z=4-5$ and under-predict at $z=6-7$
    (Figure~\ref{fig:smf}). We note, however, that both sets of observations are
    inconsistent with one another, and have, at times, large errors attached to
    them.
  \item The sSFRs of our model galaxies (Figure~\ref{fig:ssfr}) show the
    observed trend of increasing normalisation with redshift, but do not
    reproduce the observed mass dependence, indicating instead that galaxies of
    all masses the same level of activity. It is unclear as to whether this is
    caused by observational bias, or by an under-estimate of AGN feedback in the
    model.
\end{enumerate}
In summary, the {\sc L-Galaxies} model has mixed success in reproducing
observations at high redshift. It provides a reasonable match to both the
SFR\,DF and the low-mass end of the SMF, but fails to show the observed
mass-dependence of the sSFR. The predicted UV\,LF is highly-dependent upon an
ad-hoc scaling with redshift of the dust model. 

In \cite{Yates2013} we added a detailed model of the chemical enrichment in {\sc
  L-Galaxies} by adding a delayed enrichment from stellar winds and supernovae,
as well as metallicity-dependent yields and the tracking of eleven heavy
elements (including O, Mg and Fe).  This shows promising results in reproducing
the mass-metallicity relation at $z=0$, although the chemical enrichment at
high-redshift remains untested and is something we will look at in the future. That
then will provide a more realistic prediction of the metallicity of galaxies at
early times.  In future work, we will also add a physically-motivated model for
dust growth and destruction, and consider the effect of extinction from the
inter-galactic medium.


\section*{Acknowledgments}

The authors would like to thank the anonymous referee for valuable comments and suggestions
that helped us to improve this paper. We would also like to thank Kenneth Duncan, Brett Salmon 
and Renske Smit for providing us with their observational data, in some cases in advance of
publication, and Chaichalit Srisawat and David Sullivan for their advice and assistance.

The authors contributed in the following way to this paper.  SJC undertook the
vast majority of the data analysis and produced the figures; he also
provided a first draft of the paper.  PAT \& SMW jointly supervised SJC and led
the interpretation of the results.  BMBH provided expertise on the
interpretation of the {\sc L-Galaxies} model.

The model data used in this paper was generated on the DiRAC Data Centric system
at Durham University, operated by the Institute for Computational Cosmology on
behalf of the STFC DiRAC HPC Facility ({\tt www.dirac.ac.uk}). This equipment was
funded by BIS National E-infrastructure capital grant ST/K00042X/1, STFC
capital grant ST/H008519/1, and STFC DiRAC Operations grant ST/K003267/1 and
Durham University.  Much of the data analysis was undertaken on the {\sc Apollo}
cluster at Sussex University.

SJC acknowledges the support of his PhD studentship from the Science and
Technology Facilities Council (STFC).  PAT \& SMW acknowledge support from the
Science and Technology Facilities Council (grant number ST/L000652/1). BMBH was
supported by Advanced Grant 246797 ``GALFORMOD'' from the European Research
Council.

\bibliographystyle{mn2e}

\bibliography{uvlf}


\bsp

\label{lastpage}

\end{document}